\documentclass{article}

\usepackage{a4wide}
\usepackage{amsmath}
\usepackage{amsfonts}
\usepackage{amssymb}
\usepackage{latexsym}
\usepackage{ifthen}
\usepackage{theorem}
\usepackage{graphicx}
\usepackage{array}
\usepackage{caption,subcaption}

\newtheorem{prop}{Proposition} 
\newtheorem{lem}[prop]{Lemma}
\newtheorem{theo}[prop]{Theorem}
\newtheorem{cor}[prop]{Corollary}

\newcommand{\pf}{\noindent{\em Proof: }}
\newcommand{\qed}{\hfill \(\blacksquare\)\\}

\title{Phylogenetic consensus networks: Computing a consensus of 1-nested phylogenetic networks}

\date{\today}

\author{Katharina Huber and Vincent Moulton\\University of East Anglia, School of Computing Sciences\\ Norwich, NR4 7TJ, UK\\\{k.huber,v.moulton\}@uea.ac.uk
	\and Andreas Spillner\\Merseburg University of Applied Sciences\\ 06217 Merseburg, Germany\\andreas.spillner@hs-merseburg.de}

\begin{document}
	
	\date{\today}
	
	\maketitle

\begin{abstract}	
An important and well-studied problem in phylogenetics 
is to compute a \emph{consensus tree} so as to summarize the common features within 
a collection of rooted phylogenetic trees, all whose leaf-sets are bijectively labeled by the 
same set~\(X\) of species.
More recently, however, it has become of interest to find a consensus 
for a collection of more general, rooted directed acyclic graphs all of whose 
sink-sets are bijectively labeled by~\(X\), so called rooted \emph{phylogenetic networks}.
These networks are used to analyse  the evolution of species that cross with one another, 
such as plants and viruses. In this paper, we introduce
an algorithm for computing a consensus for a collection of 
so-called 1-\emph{nested} phylogenetic networks.
Our approach builds on a previous result by Rosell\'o et al.
that describes an encoding for any 1-nested phylogenetic network 
in terms of a collection of ordered pairs of subsets of~\(X\).
More specifically, we characterize those collections of ordered pairs that 
arise as the encoding of some 1-nested phylogenetic network, and then use this 
characterization to 
compute a \emph{consensus network} 
for a collection of~$t$ 1-nested networks in $O(t|X|^2+|X|^3)$ time.
Applying our algorithm to a collection of phylogenetic trees
yields the well-known majority rule consensus tree.
Our approach leads to several new directions for future work, and we 
expect that it should provide a useful new tool to 
help understand complex evolutionary scenarios.
\end{abstract}

\section{Introduction}
\label{section:introduction}

\begin{figure}[t]
\centering
\includegraphics[scale=0.9]{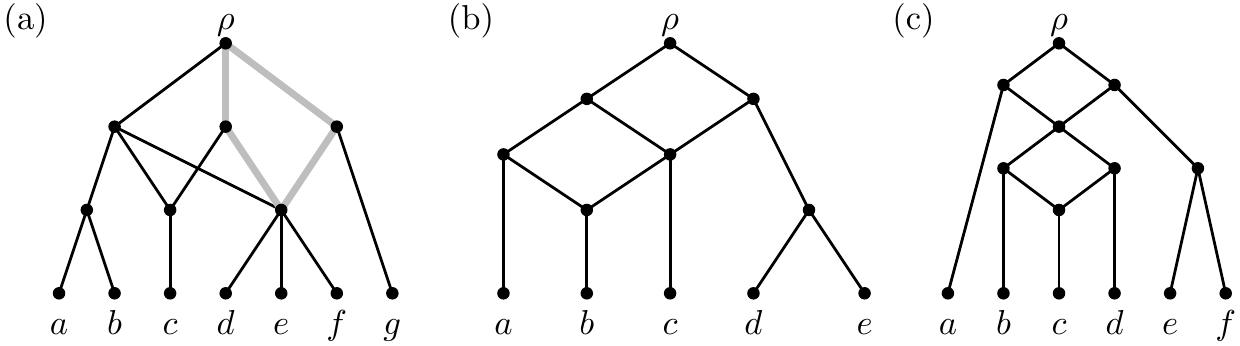}
\caption{(a) A phylogenetic network with root~\(\rho\) on \(X = \{a,b,c,d,e,f,g\}\). Arcs \((u,v)\)
are always drawn with \(u\) above \(v\). The grey arcs indicate a reticulation cycle.
(b) A 2-hybrid network on \(X = \{a,b,c,d,e\}\).
(c) A 2-hybrid, 1-nested network on \(X=\{a,b,c,d,e,f\}\).}
\label{figure:networks:introduction}
\end{figure}

In recent years, phylogenetic networks have become an important tool
for analyzing the evolution of species, and their study is 
an active area in phylogenetics~\cite{elworth2019advances,huson2010phylogenetic}. 
Given a finite non-empty set~$X$ of species, a (rooted) \emph{phylogenetic network}
on~$X$ is a directed acyclic
graph with a single source vertex~\(\rho\) (called the \emph{root})
whose set  of sinks (also called \emph{leaves}) are in bijective
correspondence with the species in~$X$ (see e.g. Figure~\ref{figure:networks:introduction}(a)).
Note that it is usually assumed  that such networks do not contain vertices whose indegree and 
outdegree are both~1.  Phylogenetic networks generalize 
(rooted) \emph{phylogenetic trees}, networks in which every vertex has indegree at most~1,
and they are particularly useful in studying the evolution 
of species that cross with one another, for
example plants or viruses~\cite{morrison2005networks}.

A well-studied class of phylogenetic networks is the class of
\emph{2-hybrid, 1-nested networks}~\cite{rossello2009all}
which are defined as follows. A phylogenetic network is
\emph{2-hybrid} if every vertex has indegree at most~2
(see e.g. Figure~\ref{figure:networks:introduction}(b)).
A \emph{reticulation cycle} in a phylogenetic network consists of two directed paths 
that have the same start vertex and the same end vertex
but no other vertices in common. A 2-hybrid phylogenetic network is \emph{1-nested} 
if no pair of reticulation cycles have an arc in common 
(see e.g. Figure~\ref{figure:networks:introduction}(c)). Important
subclasses of 2-hybrid, 1-nested networks include \emph{galled trees}
(in which no pair of reticulation cycles have a vertex in common~\cite{gusfield2003efficient}) and 
\emph{level-1 networks} (in which every reticulation cycle 
contains only one vertex with indegree~2~\cite{choy2005computing}).
In the rest of this paper, we refer to 2-hybrid, 1-nested phylogenetic networks 
simply as \emph{1-nested networks}.  
Various software packages are used to compute 
1-nested networks from biological data-sets including Dendroscope~\cite{huson2012dendroscope},
Lev1athan~\cite{huber2010practical} and Trilonet~\cite{oldman2016trilonet}.

Since alternative 1-nested networks may result for 
a data-set depending on which software is used to compute them, it is 
of interest to  develop new approaches to find a consensus
for  a collection~\(\mathfrak{C}\) of 1-nested networks in the form of a single 1-nested network.
The overarching aim is that this \emph{consensus network} should exhibit structures 
that are shared by many of the networks in~\(\mathfrak{C}\).
Note that the more specific problem of finding a consensus for a collection $\mathfrak{C}$ of
phylogenetic trees on~$X$ has been considered in phylogenetics
for many years (see~\cite{bryant2003classification} for 
a comprehensive review), and it is also well-studied in
classification theory (see~\cite{leclerc1998consensus} for a review).
One of the most popular consensus methods used for phylogenetic trees, 
is the {\em majority rule}~\cite{margush1981consensusn} approach, which we now recall.

\begin{figure}[t]
\centering
\includegraphics[scale=0.9]{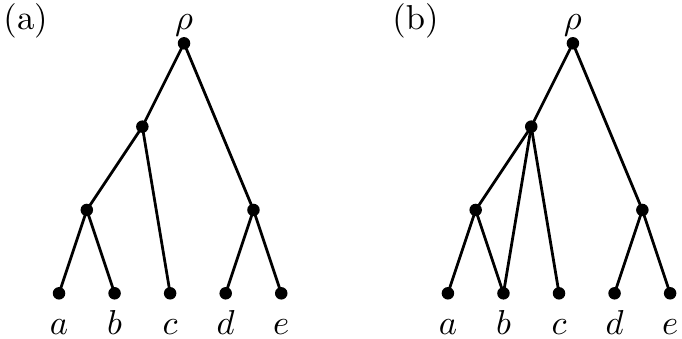}
\caption{(a) A phylogenetic tree \(\mathcal{T}\) on \(X=\{a,b,c,d,e\}\) that induces the set
\(\mathcal{C}(\mathcal{T}) = \{\{x\}:x \in X\} \cup \{\{a,b\},\{a,b,c\},\{d,e\},X\}\) of clusters.
(b)~A 1-nested network \(\mathcal{N}\) on \(X\)
that induces the same set 
of clusters as the phylogenetic tree~\(\mathcal{T}\), i.e. \(\mathcal{C}(\mathcal{N}) = \mathcal{C}(\mathcal{T})\). The set pair system induced by \(\mathcal{N}\) is
\(\theta(\mathcal{N}) = \{(\{x\},\emptyset) : x \in X\} \cup \{(\{a\},\{b\}),(\{a,b,c\},\emptyset),(\{d,e\},\emptyset),(X,\emptyset)\}\).
This set pair system encodes~\(\mathcal{N}\) and it differs from the set pair system
\(\theta(\mathcal{T}) = \{(A,\emptyset) : A \in \mathcal{C}(\mathcal{T})\}\) induced by~\(\mathcal{T}\).}
\label{figure:clusters:vs:set:pairs}
\end{figure}

First, each tree $\mathcal{T} \in \mathfrak{C}$ is broken down into the 
set \(\mathcal{C}(\mathcal{T})\) of \emph{clusters} that it induces
on the set~\(X\) (i.e. the collection of subsets of~$X$, one
subset~\(C(u)\) for each vertex $u$ in~\(\mathcal{T}\),
such that \(C(u)\) contains those $x \in X$
that can be reached from $u$ by a directed path in~\(\mathcal{T}\); 
see Figure~\ref{figure:clusters:vs:set:pairs}(a)).
Then those clusters in~\(\mathcal{C}(\mathcal{T})\) that are
induced by more than half of the trees in $\mathfrak{C}$ are kept. It can be shown 
that the resulting set of clusters uniquely defines, or \emph{encodes}, a phylogenetic tree on~\(X\).
The phylogenetic tree obtained in this way is called
the \emph{majority rule consensus tree} of~$\mathfrak{C}$.
Note that the majority rule approach has been extended to {\em unrooted} phylogenetic networks 
(see e.g.~\cite{holland2004using}); in contrast, the problem of finding
a consensus for a collection of (rooted) phylogenetic networks 
remains relatively unexplored (see~\cite{huson2012dendroscope}
where some approaches are mentioned). 

In this paper, we shall generalize the majority rule
method to 1-nested networks and, in this way, obtain a
\emph{consensus network} for any collection of such networks.
We now briefly outline our approach.
First note that the definition of the set \(\mathcal{C}(\mathcal{T})\)
of clusters induced by a phylogenetic tree~\(\mathcal{T}\) can also
be applied more generally to phylogenetic networks \(\mathcal{N}\),
and we denote by \(\mathcal{C}(\mathcal{N})\) the set of clusters
induced by \(\mathcal{N}\). In general, however,
the set \(\mathcal{C}(\mathcal{N})\) does not encode~\(\mathcal{N}\)
(see~\cite{gambette2012encodings,gambette2017challenge}).
Therefore, we consider {\em set pairs on $X$} instead of clusters.
Set pairs are ordered pairs~\((S,H)\) of subsets of \(X\) with 
\(S \neq \emptyset\) and \(S \cap H = \emptyset\). Each vertex~\(u\)
in a phylogenetic network~\(\mathcal{N}\) on~\(X\) induces such a set pair
by putting~\(S\) to be the set of those elements in the cluster \(C(u)\)
that can be reached
from the root of \(\mathcal{N}\) only by directed paths that contain~\(u\)
and putting \(H = C(u) \setminus S\) (see Figure~\ref{figure:clusters:vs:set:pairs}(b)). 
It follows from~\cite[Corollary 5]{CL11} that the equivalence class
of every 1-nested network~\(\mathcal{N}\) (with respect to a
natural equivalence relation on phylogenetic networks described
in Section~\ref{section:preliminaries}) is encoded by the set
\(\theta(\mathcal{N})\) of set pairs induced by~\(\mathcal{N}\).

Here we shall take this result 
a step further and characterize those sets of set pairs, or \emph{set pair systems}, that 
are induced by 1-nested networks (see Theorem~\ref{char:theo:encodings:nc}). 
Once we have this characterization,
we then leverage it to compute a consensus of a collection of 1-nested networks
using a similar strategy to the majority rule approach for
phylogenetic trees. In particular, for $t \ge 1$, we prove that for a 
collection of~$t$ 1-nested networks, all on the same set~\(X\) with~\(n\) elements, an 
analogue of the majority rule consensus tree can be computed in $O(tn^2+n^3)$ time 
(see Theorem~\ref{theorem:computing:consensus}). Note
that in case all of the 1-nested networks in the input collection
are phylogenetic trees then our approach will generate the
majority rule consensus tree.

The rest of the paper is organized as follows. In Section~\ref{section:preliminaries} 
we describe the above-mentioned natural equivalence
relation on 1-nested networks,
and show that we can encode any resulting equivalence class in
terms of a set pair system.
In Section~\ref{section:poset:pairs:of:sets}, we first present
some more notation related to set pair systems
and then introduce a special class of
such systems called 1-nested compatible set pair systems.
In Section~\ref{section:characterization:encodings}, we show that 
these 1-nested compatible set pair systems
are precisely those set pair systems which are induced
by 1-nested networks. In Section~\ref{section:computing:consensus}, we show
how to compute a consensus for a collection of 1-nested networks.
We conclude with a list of open problems in Section~\ref{section:discussion}. 

\section{Encoding compressed 1-nested networks}
\label{section:preliminaries}

\begin{figure}
\centering
\includegraphics[scale=0.85]{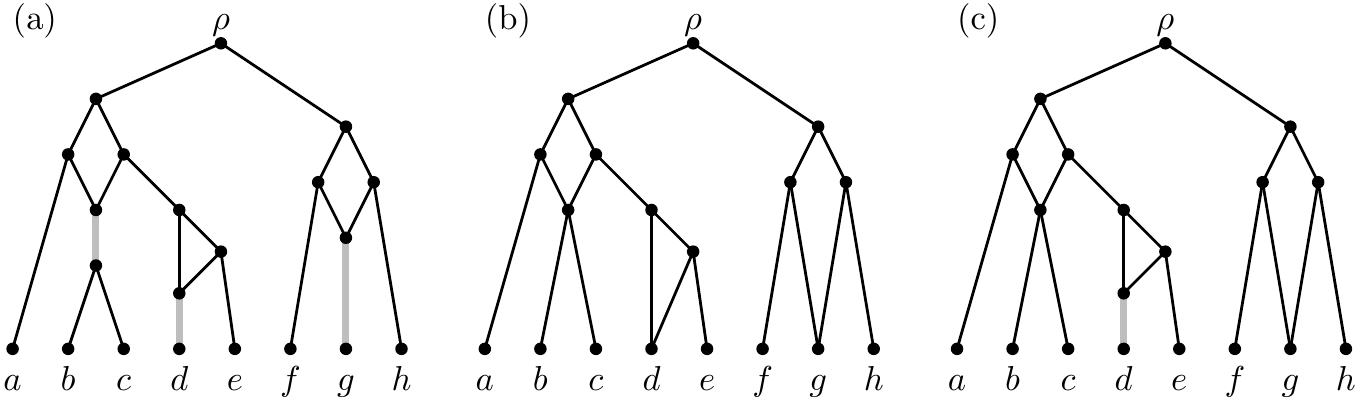}
\caption{Three 1-nested networks on \(X=\{a,b,c,d,e,f,g,h\}\) that are members of the same equivalence class.
(a) The fully expanded network in the equivalence class which does not contain any vertices with both
indegree and outdegree at least~2. (b) The compressed network in the equivalence class that is obtained
from the fully expanded network by collapsing the grey arcs. (c) A network in the
equivalence class that is neither fully expanded nor compressed.}
\label{figure:compressed:networks}
\end{figure}

In this section, we introduce compressed 1-nested networks, which represent
equivalence classes of 1-nested networks. From a biological point of view,
all 1-nested networks in such an equivalence class describe the same flow
of genetic information from the root of the network to the species at its
leaves (see Figure~\ref{figure:compressed:networks}).
Mathematically, it is more convenient to work with
compressed 1-nested networks as they are directly encoded by their induced set pairs.
To make this and the terms used informally
in the introduction more precise, we begin
by recalling some standard graph theory terminology.

A \emph{directed graph} \(N=(V,A)\) consists of a finite non-empty set
\(V\) and a subset \(A \subseteq V \times V\). The elements of
\(V\) and \(A\) are referred to as \emph{vertices} and \emph{arcs}
of~\(N\), respectively. A directed graph~\(N\) is \emph{acyclic} if there is
no directed cycle in~\(N\). Moreover, a directed acyclic graph (DAG)
\(N\)~is \emph{rooted} if there exists
a vertex \(\rho \in V\) with indegree~\(0\), called the \emph{root} of \(N\),
such that for every \(u \in V\) there is a directed path from \(\rho\) to \(u\).
In a rooted DAG, a \emph{leaf} is a vertex with outdegree~0,
a \emph{tree vertex} is a vertex with indegree at most~1 and
a \emph{reticulation vertex} is a vertex with indegree at least~2.
Note that the root of a rooted DAG is considered a tree vertex.
Moreover, in a rooted DAG~\(N\), we call a vertex~\(u\) a \emph{child}
of a vertex~\(v\) and, similarly, \(v\) the \emph{parent} of \(u\)
if \((v,u)\) is an arc of \(N\).
A \emph{reticulation cycle} \(\mathcal{C} = \{P,P'\}\) 
in a rooted DAG consists of two directed paths \(P\) and \(P'\)
such that \(P\) and \(P'\) have the same start vertex and the same end vertex
but no other vertices in common.

From now on, $X$ will denote a finite, non-empty set.
A \emph{compressed 1-nested network} \(\mathcal{N} = ((V,A),\varphi)\) on \(X\)
is a rooted DAG \(N=(V,A)\) together with a bijective map \(\varphi\) 
from \(X\) to the set of leaves of \(N\) such that:
\begin{itemize}
\setlength{\itemindent}{5pt}
\item[(i)]
No vertex of \(N\) has outdegree~1.
\item[(ii)]
All vertices of \(N\) have indegree at most~2.
\item[(iii)]
No two distinct reticulation cycles in \(N\) have an arc in common.
\end{itemize}
Note that general 1-nested networks may contain arcs \((u,v)\) such that
\(u\) has indegree~2 and outdegree~1 and \(v\) has indegree~1.
In Figure~\ref{figure:compressed:networks} these types of arcs are
highlighted in grey. Such arcs do not have any impact in the flow of genetic
information from the root of the network to its leaves and induce a
natural equivalence relation on 1-nested networks (see also~\cite[p.251]{steel2016phylogeny}
for the concept of compression in more general phylogenetic networks). For our purposes,
it will be convenient to work with that member of the equivalence class that
does not contain any such arcs, that is, we restrict to precisely the
compressed 1-nested networks defined above.

We next describe an encoding of compressed 1-nested networks.
A vertex $u$ in a rooted DAG~\(N\) is a \emph{descendant} of a vertex $v$
if there exists a directed path (possibly of length zero) 
from the root of \(N\) to \(u\) that contains \(v\). 
A descendant \(u\) of \(v\) is a \emph{strict descendant} if
every path from the root to $u$ contains $v$. Otherwise \(u\) is called
a \emph{non-strict descendant} of \(v\).
Now, given a compressed 1-nested network $\mathcal{N} = ((V,A),\varphi)$ on \(X\) and a vertex $u \in V$,
define \(C(u)\) to be the set of those \(x \in X\) with \(\varphi(x)\) a descendant of $u$,
$S(u)$ to be the set of those \(x \in X\) with \(\varphi(x)\) a strict descendant of $u$ and $H(u)$
the set of those \(x \in X\) with \(\varphi(x)\) a non-strict descendant of $u$ in $X$.
In~\cite{moret2004phylogenetic} the ordered 3-tuple
\((S(u),H(u),X \setminus C(u))\) was introduced as the so-called \emph{tripartition}
associated with vertex~\(u\). In view of the redundancy of the information stored in
the tripartition we will focus on the first two components and denote them
by $\theta(u)=(S(u),H(u))$.
Note that \(S(u) \cap H(u) = \emptyset\) for every vertex \(u\) of \(\mathcal{N}\).
Also note that, for every vertex \(u\),
the set $S(u)$ is always non-empty while $H(u)$ may be empty (see~\cite[p.~416]{CL11}).
In addition we have the following property.

\begin{lem}
\label{lemma:one:nested:set}
Suppose \(\mathcal{N} = ((V,A),\varphi)\) is a compressed 1-nested network on \(X\).
Then, for any two distinct vertices \(u,v \in V\), we have \(\theta(u) \neq \theta(v)\).
\end{lem}

\pf
Let \(u\) and \(v\) be two distinct vertices of \(\mathcal{N}\).
First it can be checked that if \(u\) and \(v\) are both contained
in a single reticulation cycle then we must have \(\theta(u) \neq \theta(v)\).

So assume that \(u\) and \(v\) are not contained in a single reticulation cycle.
If there exists a directed path \(P\) starting from the root \(\rho\) of \(\mathcal{N}\) that contains
\(u\) and \(v\) (assuming without loss of generality that \(u\) comes before \(v\) on \(P\)) 
it can be checked that we must have either \(S(v) \cup H(v) \subseteq S(u)\) or
\(S(v) \cup H(v) \subseteq H(u)\). Assume for contradiction that
\(\theta(u) =\theta(v)\).

In view of \(S(v) \neq \emptyset\) and \(S(v) \cap H(v) = \emptyset\) this is only
possible if \(H(v) = H(u) = \emptyset\). Hence, \(u\) must be a vertex 
with outdegree~1 and \((u,v)\) is an arc in \(\mathcal{N}\), in contradiction
to the fact that \(\mathcal{N}\) is a compressed 1-nested network. 

Now consider the situation where there is no directed path starting from the root of \(\mathcal{N}\)
that contains both \(u\) and \(v\). It can be checked that this implies
\((S(u) \cup H(u)) \cap (S(v) \cup H(v)) = \emptyset\) and, thus, 
\(\theta(u) \neq \theta(v)\).
\qed

Compressed 1-nested networks \(\mathcal{N}_1=((V_1,A_1),\varphi_1)\) 
and \(\mathcal{N}_2=((V_2,A_2),\varphi_2)\) on \(X\) are \emph{isomorphic}
if there exists a DAG-isomorphism \(f:V_1 \rightarrow V_2\) such that \(f(\varphi_1(x)) = \varphi_2(x)\), for all \(x \in X\).
Defining \(\theta(\mathcal{N}) = \{\theta(u) : u \in V\}\) for any compressed 
1-nested network \(\mathcal{N} = ((V,A),\varphi)\) on \(X\),
the following is a consequence of~\cite[Cor.~5]{CL11} and 
Lemma~\ref{lemma:one:nested:set}.

\begin{theo}
\label{theorem:theta:is:encoding}
Suppose that  \(\mathcal{N}_1\) 
and \(\mathcal{N}_2\) are compressed 1-nested networks.
Then \(\theta(\mathcal{N}_1) = \theta(\mathcal{N}_2)\) if and only if 
$\mathcal{N}_1$ and $\mathcal{N}_2$ are isomorphic.
\end{theo}

In view of Theorem~\ref{theorem:theta:is:encoding} the set \(\theta(\mathcal{N})\) can be viewed as
an \emph{encoding} of the isomorphism class of~\(\mathcal{N}\), for any compressed 1-nested network~\(\mathcal{N}\).

\section{Set pair systems}
\label{section:poset:pairs:of:sets}

In Section~\ref{section:preliminaries}, we have 
associated to any compressed 1-nested network \(\mathcal{N}\) on~\(X\)
an encoding in the form of the set \(\theta(\mathcal{N})\),
i.e. a non-empty collection \(\mathcal{S}\)
of ordered pairs \((S,H)\) of subsets of \(X\) with \(S \neq \emptyset\)
and \(S \cap H = \emptyset\). We call a collection \(\mathcal{S}\) 
of ordered pairs of subsets of \(X\) with these latter two properties
a \emph{set pair system} on \(X\).
In this section, we give a list of properties that a set pair system 
arising from a compressed 1-nested network on~\(X\) must necessarily satisfy.
In Section~\ref{section:characterization:encodings}, we will then
show that this list of properties actually characterizes set pair systems that are encodings of
isomorphism classes of 1-nested networks.

As a first step towards giving this characterization,
we define, for any set pair system \(\mathcal{S}\) on \(X\), a binary relation \(<\) on~\(\mathcal{S}\)
by putting \((S_1,H_1) < (S_2,H_2)\) for two distinct \((S_1,H_1),(S_2,H_2) \in \mathcal{S}\) if one the
following holds:
\begin{itemize}
\setlength{\itemindent}{3pt}
\item[(a)]
\(S_1 \cup H_1 \subseteq S_2\)
\item[(b)]
\(S_1 \cup H_1 \subseteq H_2\)
\item[(c)]
\(S_1 \subsetneq S_2\) and \(H_1 = H_2 \neq \emptyset\)
\end{itemize}
Note that conditions (a)-(c) are mutually exclusive.
In addition, we write \((S_1,H_1) \leq (S_2,H_2)\) if \((S_1,H_1) < (S_2,H_2)\) or \((S_1,H_1) = (S_2,H_2)\).

\begin{lem}
\label{lemma:is:partial:order}
The binary relation \(\leq\) is a partial ordering
for every set pair system \(\mathcal{S}\) on \(X\).
\end{lem}

\pf
Let \(\mathcal{S}\) be a set pair system on \(X\).
The relation \(\leq\) on \(\mathcal{S}\) is reflexive by definition.
To establish that \(\leq\) is also antisymmetric, consider 
\((S_1,H_1)\), \((S_2,H_2) \in \mathcal{S}\) with \((S_1,H_1) \leq (S_2,H_2)\) and
\((S_2,H_2) \leq (S_1,H_1)\). Assume for contradiction that
\((S_2,H_2) \neq (S_1,H_1)\). Then, by the definition of the binary relation \(<\),
precisely one condition from each of the two following columns must hold:
\begin{align*}
\bullet&S_1 \cup H_1 \subseteq S_2 & \bullet&S_2 \cup H_2 \subseteq S_1\\
\bullet&S_1 \cup H_1 \subseteq H_2 & \bullet&S_2 \cup H_2 \subseteq H_1\\
\bullet&S_1 \subsetneq S_2 \ \text{and} \ H_1 = H_2 \neq \emptyset & \bullet&S_2 \subsetneq S_1 \ \text{and} \ H_1 = H_2 \neq \emptyset
\end{align*}
It can be checked that every combination of two
conditions yields a contradiction, as required.

It remains to show that \(\leq\) is transitive. So, consider three
pairs \((S_1,H_1)\), \((S_2,H_2)\), \((S_3,H_3) \in \mathcal{S}\) with
\((S_1,H_1) \leq (S_2,H_2)\) and \((S_2,H_2) \leq (S_3,H_3)\).
Note that \((S_1,H_1) = (S_2,H_2)\) or \((S_2,H_2) = (S_3,H_3)\)
immediately implies \((S_1,H_1) \leq (S_3,H_3)\). Therefore, it remains
to consider \((S_1,H_1) < (S_2,H_2)\) and \((S_2,H_2) < (S_3,H_3)\).
Then, by the definition of $<$, precisely one condition
from each of the columns above must hold with the
index 1 replaced by 3 in the right column. By checking
every combination of two conditions, it follows that
$(S_1,H_1)<(S_3,H_3)$, as required.
\qed

Next we present properties that set pair systems arising from compressed 1-nested networks
must satisfy (see Proposition~\ref{proposition:network:set:pair:systems}).
More specifically, we call a set pair system~\(\mathcal{S}\) on~\(X\) \emph{1-nested compatible} if
it has the following properties:
\begin{itemize}
\setlength{\itemindent}{16pt}
\item[(NC1)]
\((X,\emptyset) \in \mathcal{S}\).
\item[(NC2)]
\((\{x\},\emptyset) \in \mathcal{S}\), for all \(x \in X\).
\item[(NC3)]
\((S,H) \in \mathcal{S}\) with \(H \neq \emptyset\) implies \((H,\emptyset) \in \mathcal{S}\).
\item[(NC4)]
For any two distinct \((S_1,H_1)\), \((S_2,H_2) \in \mathcal{S}\) precisely one of
\begin{align*}
\bullet &(S_1,H_1) < (S_2,H_2)\\
\bullet &(S_2,H_2) < (S_1,H_1)\\
\bullet &(S_1 \cup H_1) \cap (S_2 \cup H_2) = \emptyset\\
\bullet &S_1 \cap S_2 = \emptyset \ \text{and} \ H_1 = H_2 \neq \emptyset
\end{align*}
holds.
\item[(NC5)]
There are no three \((S_1,H_1)\), \((S_2,H_2)\), \((S_3,H_3) \in \mathcal{S}\) with
\(H_1 = H_2 = H_3 \neq \emptyset\), \(S_1 \cap S_2 = \emptyset\) and
either \(S_1 \cup S_2 \subseteq S_3\) or \((S_1 \cup S_2) \cap S_3 = \emptyset\).
\end{itemize}

It can be checked with the set pair systems given below,
that Properties~(NC1)-(NC5) are independent of one another 
in the sense that for
every \(i \in \{1,2,3,4,5\}\) there exists a set pair system \(\mathcal{S}_i\) on some set $X$ that satisfies
all of these properties except for property (NC$i$):
\begin{align*}
\mathcal{S}_1 &= \{(\{a\},\emptyset),(\{b\},\emptyset)\} \ \text{on} \ X=\{a,b\}\\
\mathcal{S}_2 &= \{(X,\emptyset)\} \ \text{on} \ X=\{a,b\}\\
\mathcal{S}_3 &= \{(X,\emptyset),(\{a\},\emptyset),(\{b\},\emptyset),(\{c\},\emptyset),(\{a\},\{b,c\})\} \ \text{on} \ X=\{a,b,c\}\\
\mathcal{S}_4 &= \{(X,\emptyset),(\{a\},\emptyset),(\{b\},\emptyset),(\{c\},\emptyset),(\{a,b\},\{c\}),(\{a,c\},\{b\})\}\\
 &\qquad \text{on} \ X=\{a,b,c\}\\
\mathcal{S}_5 &= \{(X,\emptyset),(\{a\},\emptyset),(\{b\},\emptyset),(\{c\},\emptyset),(\{d\},\emptyset),(\{a\},\{d\}),(\{b\},\{d\}),(\{c\},\{d\})\}\\
 &\qquad \text{on} \ X=\{a,b,c,d\}
\end{align*}
In view of our aim to compute a consensus
of a collection of compressed 1-nested networks,
a key aspect of properties (NC1)-(NC5)  is that they can be checked
locally for any set pair system~\(\mathcal{S}\), that is, by inspecting only
subsets of \(\mathcal{S}\) of small constant size. 

\begin{prop}
\label{proposition:network:set:pair:systems}
For any compressed 1-nested network \(\mathcal{N}\) on \(X\) the set pair system \(\theta(\mathcal{N})\)
is 1-nested compatible.
\end{prop}

\pf
Let \(\mathcal{N} = ((V,A),\varphi)\) be a compressed 1-nested network on \(X\) with root~\(\rho\). Then we have
\(\theta(\rho)= (X,\emptyset) \in \theta(\mathcal{N})\), implying (NC1). Moreover,
for every \(x \in X\), the vertex \(\varphi(x)\) is a leaf of \(\mathcal{N}\)
and we have \(\theta(\phi(x)) = (\{x\},\emptyset) \in \theta(\mathcal{N})\), implying (NC2).

Next consider a vertex \(v \in V\) such that \(H(v) \neq \emptyset\).
Then there exists a unique reticulation cycle \(\mathcal{C} = \{P,P'\}\) in \(\mathcal{N}\)
such that \(v\) is a vertex on the directed path \(P\). Note 
that since $H(v)\not=\emptyset$ and $\mathcal N$ is 1-nested, $v$ cannot be the start or
end vertex of $P$. Let \(u \neq v\) denote the end vertex of \(P\).
Then \(\theta(u) = (H(v),\emptyset) \in \theta(\mathcal{N})\), implying (NC3).
 
To establish (NC4), consider two distinct vertices \(u,v \in V\). In view of Lemma~\ref{lemma:one:nested:set}
we must have $\theta(u)\not=\theta(v)$.
First we consider the case that \(u\) and \(v\) are both vertices in some reticulation cycle \(\mathcal{C}=\{P,P'\}\).
This can lead to the following configurations (ignoring symmetric configurations obtained by switching the roles of \(P\) and \(P'\)):
\begin{itemize}
\item
\(u\) is the start vertex of \(P\) and \(v\) is another vertex on \(P\).
Then we have \(S(v) \cup H(v) \subseteq S(u)\) implying \((S(v),H(v)) < (S(u),H(u))\), as required.
\item
\(u\) is a vertex of \(P\), but neither its start nor its end vertex, and \(v\) is the end vertex of \(P\).
Then we have \(S(v) = H(u)\) and \(H(v) = \emptyset\) implying \((S(v),H(v)) < (S(u),H(u))\), as required.
\item
\(u\) and \(v\) are both vertices on \(P\) with \(u\) coming before \(v\) and both
vertices are neither the start nor the end vertex of \(P\).
Then we have \(S(v) \subsetneq S(u)\) and \(H(v) = H(u) \neq \emptyset\) implying \((S(v),H(v)) < (S(u),H(u))\), as required.
\item
\(u\) is a vertex on \(P\) and \(v\) is a vertex on \(P'\) but both vertices are neither the start nor the end vertex
of \(P\) and \(P'\), respectively.
Then we have \(S(u) \cap S(v) = \emptyset\) and \(H(u) = H(v) \neq \emptyset\), as required.
\end{itemize}
Next we consider the case that \(u\) and \(v\) are not contained in the same reticulation cycle.
This can lead to the following configurations:
\begin{itemize}
\item
There is a directed path \(P\) in \(\mathcal{N}\) starting at the root \(\rho\) that contains both \(u\) and \(v\).
Then, assuming without loss of generality that \(v\) comes before \(u\) on \(P\),
we have \(S(u) \cup H(u) \subseteq S(v)\) or \(S(u) \cup H(u) \subseteq H(v)\)
implying \((S(u),H(u)) < (S(v),H(v))\), as required.
\item
There is no directed path from the root \(\rho\) that contains both \(u\) and \(v\).
Then we have \((S(u) \cup H(u)) \cap (S(v) \cup H(v)) = \emptyset\), as required. 
\end{itemize}
Hence, \(\theta(\mathcal{N})\) satisfies (NC4).

Next, to establish (NC5), consider three distinct vertices \(u,v,w \in V\) with
\(H(u) = H(v) = H(w) \neq \emptyset\) and \(S(u) \cap S(v) = \emptyset\). Since $\mathcal N$ is 1-nested, this is only possible
if \(u,v,w\) are all vertices in the same reticulation cycle
\(\mathcal{C}=\{P,P'\}\) but none of them can be the start or end vertex
of the directed paths \(P\) and \(P'\). Since $S(u)\cap S(v)=\emptyset$,
$u$ and $v$ cannot lie on the same directed path in $\mathcal C$.
Without loss of generality,
we may therefore assume that \(u\) and \(w\) are vertices on \(P\).
Again in view of \(S(u) \cap S(v) = \emptyset\), \(v\) must be a vertex on \(P'\)
implying that also \(S(w) \cap S(v) = \emptyset\).
From this it follows that we cannot have \(S(v) \cup S(u) \subseteq S(w)\).
Moreover, assuming without loss of generality that \(u\) comes before \(w\) on \(P\),
we have \(\emptyset \not= S(w) \subsetneq S(u)\),
implying that we cannot have \((S(v) \cup S(u)) \cap S(w) = \emptyset\).~\qed

\section{1-nested compatible set pair systems are encodings}
\label{section:characterization:encodings}

In this section we prove the following result. 

\begin{theo}
\label{char:theo:encodings:nc}
Given a set pair system \(\mathcal{S}\) on \(X\) there exists a
compressed 1-nested network \(\mathcal{N}\) on \(X\) with \(\mathcal{S} = \theta(\mathcal{N})\)
if and only if \(\mathcal{S}\) is 1-nested compatible. Moreover,
if it exists then \(\mathcal{N}\) is
unique up to isomorphism.
\end{theo}

Note that, in view of
Proposition~\ref{proposition:network:set:pair:systems}, there remains
only one implication to be established to prove Theorem~\ref{char:theo:encodings:nc}. Also note that 
Theorem~\ref{char:theo:encodings:nc} is a 
generalization of the so-called ``Cluster Equivalence Theorem"
for rooted trees and hierarchies (see e.g.~\cite[Proposition 2.1]{steel2016phylogeny}).
Indeed, this equivalence theorem follows from Theorem~\ref{char:theo:encodings:nc}
by considering set-pair systems \(\mathcal{S}\)
in which $H=\emptyset$ for all $(S,H) \in \mathcal{S}$.

\begin{figure}
\centering
\includegraphics[scale=0.85]{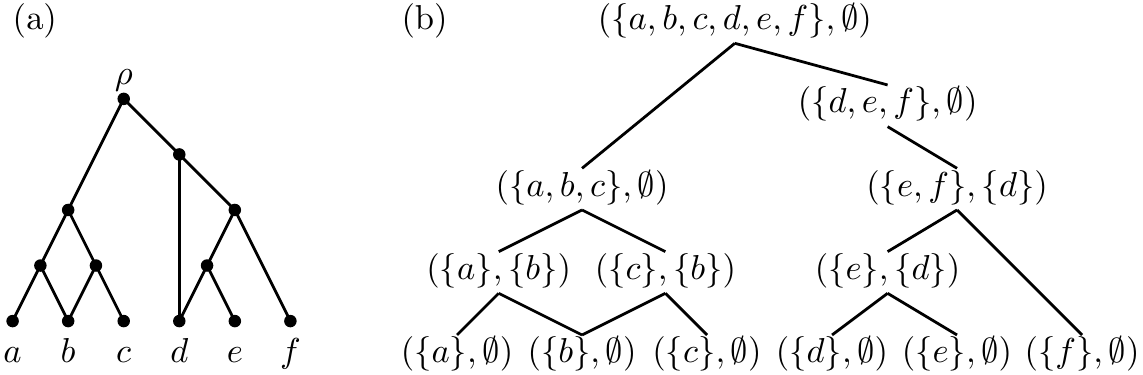}
\caption{(a) A compressed 1-nested network \(\mathcal{N}\) on \(X=\{a,b,c,d,e,f\}\).
(b) The Hasse diagram for the set pair system \(\theta(\mathcal{N})\) with respect to the 
partial ordering \(\leq\) defined in Section~\ref{section:poset:pairs:of:sets}.}
\label{figure:hasse:diagram}
\end{figure}

In our proof of Theorem~\ref{char:theo:encodings:nc}, we will use the
concept of the \emph{Hasse diagram} of a partial ordering \(\pi\) on a finite set \(M\), that is, the DAG with vertex set
\(M\) in which \((x,z) \in M \times M\) forms an arc if and only if
\(x \pi z\) holds and there is no \(y \in M \setminus \{x,z\}\) with \(x \pi y\) and \(y \pi z\). 
Our proof of Theorem~\ref{char:theo:encodings:nc}
will follow a similar strategy to that used in the proof of~\cite[Proposition 2.1]{steel2016phylogeny},
in which it is shown that, when considering the usual set inclusion as the partial ordering
on the set \(\mathcal{C}(\mathcal{T})\) of clusters induced by
a phylogenetic tree \(\mathcal{T}\), the resulting Hasse diagram is isomorphic to~\(\mathcal{T}\).
Note that, as can be seen in Figure~\ref{figure:hasse:diagram}, the Hasse diagram of the
partial ordering introduced in Section~\ref{section:poset:pairs:of:sets} on the
set pair system \(\theta(\mathcal{N})\) for a compressed 1-nested network \(\mathcal{N}\)
will, in general, not be isomorphic to \(\mathcal{N}\). More specifically, the Hasse diagram
is always missing those arcs of \(\mathcal{N}\) which occur in a directed path in a reticulation
cycle such that the path consists only of this single arc. We will come back to this
technicality in Theorem~\ref{theorem:char:ns:is:1:nested} below. 

For the rest of this section, \(\mathcal{S}\) denotes  a 1-nested compatible
set pair system on \(X\) and  \(D(\mathcal{S})\) the Hasse diagram
of the partial ordering \(\leq\) on \(\mathcal{S}\) defined in Section~\ref{section:poset:pairs:of:sets}.
The bulk of the following proof is concerned with showing that Properties~(NC1)-(NC5) suffice to establish that \(D(\mathcal{S})\) is, up to the
technicality just mentioned above, isomorphic to a compressed 1-nested network \(\mathcal{N}\)
with \(\theta(\mathcal{N}) = \mathcal{S}\).
We begin with a basic observation about \(D(\mathcal{S})\).

\begin{lem}
\label{lemma:char:hasse:diagram:root:leaves}
\(D(\mathcal{S})\) is a rooted DAG whose leaves are in one-to-one
correspondence with the elements of \(X\).
\end{lem}

\pf
By the definition of the Hasse diagram, \(D=D(\mathcal S)\) is a DAG. Moreover, in view of (NC1),
we must have \((X,\emptyset) \in \mathcal{S}\) and, by the definition
of \(\leq\), we also have \((S,H) \leq (X,\emptyset)\) for all \((S,H) \in \mathcal{S}\). Thus, \(D\) is rooted with root \((X,\emptyset)\).

Next consider an arbitrary \(x \in X\). In view of (NC2),
we have \((\{x\},\emptyset) \in \mathcal{S}\) and it follows immediately from the definition of \(\leq\) that
\((\{x\},\emptyset)\) has outdegree~0 in~\(D\). To show that the vertices of outdegree~0
in \(D\) are in one-to-one correspondence with the elements in \(X\), assume for contradiction
that there exists some \((S,H) \in \mathcal{S}\) with outdegree~0 but \((S,H) \neq (\{x\},\emptyset)\)
for all \(x \in X\). By the definition
of a set pair system we must have \(S \neq \emptyset\) and so we may select some \(x \in S\).
But then, by the definition of \(\leq\), we have \((\{x\},\emptyset) < (S,H)\), implying
that the outdegree of \((S,H)\) in \(D\) is greater than~0, a contradiction.
\qed

We next consider properties of set pairs in \(\mathcal{S}\) which give rise to vertices with 
indegree 1 in $D(\mathcal{S})$.

\begin{lem}
\label{lemma:char:indegree:one:vertices}
Suppose \((S,H) \in \mathcal{S}\) with \(H \neq \emptyset\). Then:
\begin{itemize}
\setlength{\itemindent}{5pt}
\item[(i)] $(S,H)$ has indegree~1 in~\(D(\mathcal{S})\).
\item[(ii)] For every \((S_1,H_1) \in \mathcal{S}\) with \((H,\emptyset) < (S_1,H_1) < (S,H)\),
we have \(S_1 \subsetneq S\) and \(H_1 = H\).
\item[(iii)] There exists a unique \((S_2,H_2) \in \mathcal{S}\) that is minimal with respect to \(\leq\)
such that \((S,H) < (S_2,H_2)\) and \(H_2 \neq H\). Moreover, 
\(S \cup H \subseteq S_2\) and, for every \((S_1,H_1) \in \mathcal{S}\) 
with \((H,S) < (S_1,H_1) < (S_2,H_2)\),  \(S \subsetneq S_1\) and
\(H_1 = H\).
\end{itemize}
\end{lem}

\pf
\noindent (i): Assume for contradiction that there exist two distinct \((S_1,H_1)\), \((S_2,H_2)\) in \(\mathcal{S}\)
such that \((S,H)\) is the child of both in~\(D(\mathcal{S})\).
Then, by the definition of the Hasse diagram,
we have neither \((S_1,H_1) < (S_2,H_2)\) nor \((S_2,H_2) < (S_1,H_1)\).
Hence, in view of (NC4), we must have either \((S_1 \cup H_1) \cap (S_2 \cup H_2) = \emptyset\)
or \(S_1 \cap S_2 = \emptyset\) and \(H_1 = H_2 \neq \emptyset\).
In view of \((S,H) < (S_1,H_1)\) and \((S,H) < (S_2,H_2)\),
we must have \(\emptyset \neq S \subseteq (S_1 \cup H_1) \cap (S_2 \cup H_2)\). It follows 
that \(S_1 \cap S_2 = \emptyset\) and \(H_1 = H_2 \neq \emptyset\).
But then the only way to have \((S,H) < (S_1,H_1)\) and \((S,H) < (S_2,H_2)\)
is \(S \cup H \subseteq H_1 = H_2\). From this, using the fact that 
\((H_1,\emptyset) \in \mathcal{S}\) in view of (NC3), we obtain
\((S,H) < (H_1,\emptyset) < (S_1,H_1)\), in contradiction to the fact that \((S,H)\)
is a child of \((S_1,H_1)\) in \(D(\mathcal{S})\).

\noindent (ii): By the definition of \(<\), \((H,\emptyset) < (S_1,H_1)\) implies
\(H \subseteq S_1\) or \(H \subseteq H_1\). But then we cannot have
\(S_1 \cup H_1 \subseteq S\) in view of \(S \cap H = \emptyset\).
Moreover, we cannot have \(S_1 \cup H_1 \subseteq H\) since this would
imply \((S_1,H_1) = (H,\emptyset)\) or \(S_1 = \emptyset\).
Hence, the only way to have \((S_1,H_1) < (S,H)\) is
\(S_1 \subsetneq S\) and \(H_1 = H \neq \emptyset\), as required.

\noindent (iii): First note that in view of \((X,\emptyset) \in \mathcal{S}\), \((S,H) < (X,\emptyset)\) and \(H \neq \emptyset\)
there must exist at least one \((S_2,H_2) \in \mathcal{S}\) that is minimal with respect to \(\leq\)
such that \((S,H) < (S_2,H_2)\) and \(H_2 \neq H\). 

By the definition of \(<\) and in view of \(H_2 \neq H\), we must have either \(S \cup H \subseteq S_2\) or
\(S \cup H \subseteq H_2\). Assume for contradiction that \(S \cup H \subseteq H_2\).
This implies \(H_2 \neq \emptyset\). Consider the set pair \((H_2,\emptyset)\)
which must be contained in \(\mathcal{S}\) in view of (NC3). Then we
have \((S,H) < (H_2,\emptyset) < (S_2,H_2)\) in contradiction to
\((S_2,H_2)\) being minimal. Thus, we must have \(S \cup H \subseteq S_2\), as required.

Now consider an arbitrary \((S_1,H_1) \in \mathcal{S}\) with
\((S,H) < (S_1,H_1) < (S_2,H_2)\). Since \((S_2,H_2)\) is minimal,
we must have \(H_1 = H\). Therefore, we can have
neither \(S \cup H \subseteq S_1\) in view of \(S_1 \cap H = \emptyset\)
nor \(S \cup H \subseteq H\) in view of \(S \cap H = \emptyset\) and \(S \neq \emptyset\).
Hence, by the definition of \(<\), we must have \(S \subsetneq S_1\),
as required.

Finally, assume for contradiction that there
are two distinct minimal elements \((S_2,H_2)\), \((S_2',H_2') \in \mathcal{S}\)
with \((S,H) < (S_2,H_2)\) and \(H_2 \neq H\) as well as \((S,H) < (S_2',H_2')\) and \(H_2' \neq H\).
Then there must exist some \((S_1,H) \in \mathcal{S}\) with
\((S,H) \leq (S_1,H)\), \((S_1,H) < (S_2,H_2)\) and \((S_1,H) < (S_2',H_2')\)
such that \((S_1,H)\) has indegree~2. But, in view of \(H \neq \emptyset\),
this is in contradiction to Part~(i) of this lemma.
Thus, the minimal element is unique.
\qed

We now show that every set pair in $\mathcal S$ has indegree at most 2 in \(D(\mathcal{S})\).
To do this we first prove a useful lemma concerning set pairs $(S,H) \in \mathcal S$ with $H = \emptyset$
(note that in Lemma~\ref{lemma:char:indegree:one:vertices} we considered set pairs with $H \neq \emptyset$).

\begin{lem}
\label{lemma:char:not:indegree:two:when:empty}
Suppose \((S,H) \in \mathcal{S}\) with \(H = \emptyset\). Then
$(S,H)$ has at most one parent \((S_1,H_1)\) with \(H_1 = \emptyset\) in~\(D(\mathcal{S})\), and it 
has at most two distinct parents \((S_1,H_1)\) and \((S_2,H_2)\) with
\(H_1 \neq \emptyset\) and \(H_2 \neq \emptyset\) in~\(D(\mathcal{S})\).
Moreover, if \((S,H)\) has two distinct such parents, 
then \(S_1 \cap S_2 = \emptyset\) and \(H_1 = H_2 = S\).
\end{lem}

\pf We first show that $(S,H)$ has at most one 
parent \((S_1,H_1)\) with \(H_1 = \emptyset\) in~\(D(\mathcal{S})\).
Assume for contradiction that \((S,H)\) has
two distinct parents \((S_1,\emptyset)\) and \((S_2,\emptyset)\) in~\(D(\mathcal{S})\).
Note that this implies \(\emptyset \neq S \subseteq S_1 \cap S_2\).
Moreover, it follows immediately from the definition of the Hasse diagram
that we can have neither \((S_1,\emptyset) < (S_2,\emptyset)\) nor \((S_2,\emptyset) < (S_1,\emptyset)\).
As a consequence and in view of (NC4), we have \(S_1 \cap S_2 = \emptyset\), in contradiction to
\(\emptyset \neq S \subseteq S_1 \cap S_2\).

To see that  the second statement in the lemma holds, assume for contradiction that
\((S,H)\) has at least three distinct parents \((S_1,H_1)\), \((S_2,H_2)\), \((S_3,H_3)\)
with \(H_1 \neq \emptyset\), \(H_2 \neq \emptyset\) and \(H_3 \neq \emptyset\) in~\(D(\mathcal{S})\).
Thus, by the definition of the Hasse diagram, we cannot have \((S_i,H_i) < (S_j,H_j)\) for any two distinct \(i,j \in \{1,2,3\}\).
Hence, in view of (NC4) and \(\emptyset \neq S \subseteq S_i \cup H_i\) for all \(i \in \{1,2,3\}\),
we must have \(S_1 \cap S_2 = S_1 \cap S_3 = S_2 \cap S_3 = \emptyset\) and
\(H_1 = H_2 = H_3 = S \neq \emptyset\), in contradiction to (NC5).

Finally, using the same argument, it follows that \(S_1 \cap S_2 = \emptyset\) and \(H_1 = H_2 = S\)
in case \((S,H)\) has two distinct parents \((S_1,H_1)\) and \((S_2,H_2)\) with
\(H_1 \neq \emptyset\) and \(H_2 \neq \emptyset\).
\qed

\begin{prop}
\label{prop:char:indegree:at:most:two}
Every \((S,H) \in \mathcal{S}\) has indegree at most~2 in~\(D(\mathcal{S})\).
If \((S,H)\) has two distinct parents \((S_1,H_1)\) and \((S_2,H_2)\) 
then we have \(H=\emptyset\), \(H_1 = H_2 = S\) and \(S_1 \cap S_2 = \emptyset\).
\end{prop}

\pf
First assume for contradiction that there exists some \((S,H) \in \mathcal{S}\) that
has three distinct parents \((S_1,H_1)\), \((S_2,H_2)\) and \((S_3,H_3)\) in \(D(\mathcal{S})\).
In view of Lemma~\ref{lemma:char:indegree:one:vertices}(i), this implies \(H = \emptyset\).
Moreover, in view of Lemma~\ref{lemma:char:not:indegree:two:when:empty}, we must have without loss
of generality \(H_1 = \emptyset\), \(H_2 = H_3 = S \neq \emptyset\) and \(S_2 \cap S_3 = \emptyset\).

From the definition of the Hasse diagram it follows that we 
have neither \((S_1,\emptyset) < (S_i,S)\) nor \((S_i,S) < (S_1,\emptyset)\)
for \(i \in \{2,3\}\).
Thus, we must have \(S_1 \cap (S_i \cup S) = \emptyset\) in view of (NC4).
But this is in contradiction to the fact that, in view of
\((S,\emptyset)= (S,H)<(S_1,H_1) = (S_1,\emptyset)\), we have \(\emptyset \neq S \subseteq S_1 \cap (S_i \cup S)\).
This establishes that every \((S,H) \in \mathcal{S}\) has indegree at most~2.

To finish the proof of the proposition,
assume that \((S,H) \in \mathcal{S}\) has two distinct parents \((S_1,H_1)\) and \((S_2,H_2)\)
in~\(D(\mathcal{S})\). Then, in view of Lemma~\ref{lemma:char:indegree:one:vertices}(i), we have \(H = \emptyset\).
Hence, by Lemma~\ref{lemma:char:not:indegree:two:when:empty}, we cannot have both \(H_1 = \emptyset\)
and \(H_2 = \emptyset\). Moreover, by
the same lemma,
if \(H_1 \neq \emptyset\) and \(H_2 \neq \emptyset\), we must have
\(H_1 = H_2 = S\) and \(S_1 \cap S_2 = \emptyset\), as required.

It remains to consider the case that, without loss of generality,
\(H_1 = \emptyset\) and \(H_2 \neq \emptyset\).
By the definition of the Hasse diagram, we can have
neither \((S_1,\emptyset) < (S_2,H_2)\) nor \((S_2,H_2) < (S_1,\emptyset)\).
Thus, in view of (NC4), we must have \(S_1 \cap (S_2 \cup H_2) = \emptyset\)
in contradiction to \(\emptyset \neq S \subseteq S_1 \cap (S_2 \cup H_2)\).
\qed

We now prove a lemma which will be key to understanding reticulation cycles in \(D(\mathcal{S})\).

\begin{lem}
\label{lemma:char:unique:path:nonempty}
The following statements hold:
\begin{enumerate}
	\item[(i)]
For every \((S,H) \in \mathcal{S}\) with \(H \neq \emptyset\)
there exists a unique directed path
\[P(S,H) = ((S_*,H_*),(S_k,H),(S_{k-1},H),\dots,(S_1,H),(H,\emptyset))\]
in \(D(\mathcal{S})\) with \(k \geq 1\), \((S,H) = (S_i,H)\) for some \(1 \leq i \leq k\),
\(S \cup H \subseteq S_*\) and \(H_* \neq H\).
\item[(ii)]
For any three \((S,H)\), \((S',H)\) and \((S'',H) \in \mathcal{S}\)
with \(H \neq \emptyset\), at least two of the directed paths
\(P(S,H)\), \(P(S',H)\) and \(P(S'',H)\) must coincide, 
and if \((S,H)\), \((S',H)\)  are such that the directed paths
\[P(S,H) = ((S_*,H_*),(S_k,H),(S_{k-1},H),\dots,(S_1,H),(H,\emptyset))\]
and
\[P(S',H) = ((S'_*,H'_*),(S'_l,H),(S'_{l-1},H),\dots,(S'_1,H),(H,\emptyset))\]
do not coincide, then \(S_i \cap S'_j = \emptyset\) for all \(1 \leq i \leq k\) and
\(1 \leq j \leq l\), and \((S_*,H_*) = (S'_*,H'_*)\).
\end{enumerate}
\end{lem}

\pf
\noindent(i): By (NC3) we have \((H,\emptyset) \in \mathcal{S}\).
Let \((S_*,H_*) \in \mathcal{S}\) be minimal with respect to \(\leq\)
such that \((S,H) < (S_*,H_*)\) and \(H_* \neq H\). Note that
\((S_*,H_*)\) exists and is unique by Lemma~\ref{lemma:char:indegree:one:vertices}(iii).

Now consider
\[\mathcal{S}' = \{(S',H') \in \mathcal{S}: (H,\emptyset) < (S',H') < (S,H)\}.\]
By Lemma~\ref{lemma:char:indegree:one:vertices}(ii) we have
\(S' \subsetneq S\) and \(H'=H\) for all \((S',H') \in \mathcal{S}'\).
Moreover, we must have either \(S_1' \subsetneq S_2'\) or \(S_2' \subsetneq S_1'\)
for any two distinct \((S_1',H)\), \((S_2',H) \in \mathcal{S}'\). To see this,
assume for contradiction that we have neither \(S_1' \subsetneq S_2'\) nor \(S_2' \subsetneq S_1'\).
Then, in view of (NC4), we have \(S_1' \cap S_2' = \emptyset\).
But this contradicts (NC5) for \((S_1',H)\), \((S_2',H)\) and \((S,H)\).

Next consider
\[\mathcal{S}'' = \{(S'',H'') \in \mathcal{S}: (H,S) < (S'',H'') < (S_*,H_*)\}.\]
By Lemma~\ref{lemma:char:indegree:one:vertices}(iii), we have
\(S \subsetneq S''\) and \(H'' = H\) for all \((S'',H'') \in \mathcal{S}''\).
Moreover, we must have either \(S_1'' \subsetneq S_2''\) or \(S_2'' \subsetneq S_1''\)
for any two distinct \((S_1'',H)\), \((S_2'',H) \in \mathcal{S}'\) since
otherwise \((S,H)\) or one of the elements of \(\mathcal{S}''\) has indegree~2
in contradiction to Lemma~\ref{lemma:char:indegree:one:vertices}(i).

As a consequence, we obtain a unique directed path in \(D(\mathcal{S})\) that contains \((S,H)\),
which starts at \((S_*,H_*)\), then goes through the elements of \(\mathcal{S}''\), then
through \((S,H)\), then through the elements of \(\mathcal{S}'\) and ends at \((H,\emptyset)\).

\noindent(ii):
First note that 
by Lemma~\ref{lemma:char:unique:path:nonempty}(i) all three paths \(P(S,H)\), \(P(S',H)\) and \(P(S'',H)\)
end at \((H,\emptyset)\) but, in view of Proposition~\ref{prop:char:indegree:at:most:two},
at least two of these paths must arrive at \((H,\emptyset)\) along the same
arc, implying that they coincide.

Now, suppose the paths \(P(S,H)\) and \(P(S',H)\) do not coincide.
Then, by Lemma~\ref{lemma:char:unique:path:nonempty}(i), we have
\(P(S_i,H) = P(S,H)\) for all \(1 \leq i \leq k\) and
\(P(S'_j,H) = P(S',H)\) for all \(1 \leq j \leq l\).
Since these paths are unique, we have neither
\((S_i,H) \leq (S'_j,H)\) nor \((S'_j,H) \leq (S_i,H)\)
for all \(1 \leq i \leq k\) and \(1 \leq j \leq l\) and,
thus, in view of (NC4), \(S_i \cap S'_j = \emptyset\).

Next note that, in view of \(H \neq H_*\) and 
\(\emptyset \neq H \subseteq (S_* \cup H_*) \cap (S'_l \cup H)\),
(NC4) implies \((S_*,H_*) < (S'_l,H)\) or \((S'_l,H) < (S_*,H_*)\).
By Lemma~\ref{lemma:char:indegree:one:vertices}(ii), \((H,\emptyset) < (S_*,H_*) < (S'_l,H)\)
would imply \(H_* = H\) in contradiction to \(H_* \neq H\).
Hence, we must have \((S'_l,H) < (S_*,H_*)\). By symmetry, we must
also have \((S_k,H) < (S'_*,H'_*)\). By the minimality of
\((S_*,H_*)\) and \((S'_*,H'_*)\) this implies \((S_*,H_*) \leq (S'_*,H'_*)\)
as well as \((S'_*,H'_*) \leq (S_*,H_*)\) from which we obtain
\((S_*,H_*) = (S'_*,H'_*)\) using that \(\leq\) is reflexive.
\qed

We now consider reticulation cycles in \(D(\mathcal{S})\).

\begin{prop}
\label{prop:char:reticulation:cycle:structure}
Suppose \(\mathcal{C}\) is reticulation cycle in \(D(\mathcal{S})\).
Then there exist \((S,H)\), \((S',H) \in \mathcal{S}\) with a unique \(H = H(\mathcal{C}) \neq \emptyset\)
and \(S \cap S' = \emptyset\) such that \(\mathcal{C} = \{P(S,H),P(S',H)\}\).
Moreover, if \((S'',H'') \in \mathcal{S}\) with \(H'' \neq \emptyset\) and \(H ''\neq H(\mathcal{C})\), 
then the directed path \(P(S'',H'')\) has no arc
in common with any of the directed paths in~\(\mathcal{C}\).
\end{prop}

\pf  Let \((S_0,H_0)\) be the end vertex of the directed paths in \(\mathcal{C}\).
Then \((S_0,H_0)\) has indegree~2. Let \((S,H)\) and \((S',H')\) 
denote the two parents of \((S_0,H_0)\).
Then, by Proposition~\ref{prop:char:indegree:at:most:two},
we have \(H_0 = \emptyset\) and \(H = H' = S_0\). Moreover, 
in view of Lemma~\ref{lemma:char:unique:path:nonempty},
\(\mathcal{C} = \{P(S,H),P(S',H)\}\) 
and \(S \cap S' = \emptyset\) must hold.

To see that the 
second statement in the proposition holds, assume 
for contradiction that \(P(S'',H'')\) and 
one of the directed paths in~\(\mathcal{C}\) have an arc in common.
Let \(H' = H(\mathcal{C})\) and
\[P(S'',H'') = ((S''_*,H''_*),(S''_k,H''),(S''_{k-1},H''),\dots,(S''_1,H''),(H'',\emptyset))\]
and assume that
\[P(S',H') = ((S'_*,H'_*),(S'_l,H'),(S'_{l-1},H'),\dots,(S'_1,H'),(H',\emptyset))\]
is the directed path in \(\mathcal{C}\) that has an arc in common with \(P(S'',H'')\).
Comparing the set pairs occurring in \(P(S'',H'')\) and \(P(S',H')\)
it follows that a common arc would imply
one of \(H''=H'\), \(H=\emptyset\) or \(H''=\emptyset\), a contradiction.
\qed

Now we come back to the technicality mentioned at the beginning of this section.
In particular, we modify \(D(\mathcal{S}) = (\mathcal{S},A)\) to produce a suitable phylogenetic network 
\(\mathcal{N}(\mathcal{S}) = ((\mathcal{S},A'),\varphi)\) on~\(X\) by defining
\begin{itemize}
\item
 \(\varphi:X \rightarrow \mathcal{S}\) to be the map which 
 takes \(x\) to \((\{x\},\emptyset)\) for all \(x \in X\), and
\item
\(A'\) to be the set obtained by adding 
the arc from \((S_*,H_*)\) to \((H,\emptyset)\) for every
\((S,H) \in \mathcal{S}\) with \(H \neq \emptyset\)  to \(A\)
such that the directed path
\[P(S,H) = ((S_*,H_*),(S_k,H),(S_{k-1},H),\dots,(S_1,H),(H,\emptyset))\]
given by Lemma~\ref{lemma:char:unique:path:nonempty} is not contained in a reticulation cycle in \(D(\mathcal{S})\)
(note that \((H,\emptyset)\) has indegree~1 in \(D(\mathcal{S})\)).
\end{itemize}

\begin{theo}
\label{theorem:char:ns:is:1:nested}
\(\mathcal{N}(\mathcal{S}) = ((\mathcal{S},A'),\varphi)\) is a compressed 1-nested network on \(X\).
\end{theo}

\pf
First note that the arcs added to \(D=D(\mathcal{S})\) in the construction of \(\mathcal{N}(\mathcal{S})\)
respect the partial ordering \(\leq\) and no outgoing arc is added to a leaf of \(D(\mathcal{S})\).
Therefore, Lemma~\ref{lemma:char:hasse:diagram:root:leaves} implies that
\(N=(\mathcal{S},A')\) is a rooted DAG and that \(\varphi\) is a bijection between \(X\) and the
set of leaves of \(N\).

We now show that there is no vertex in \(N\) with outdegree~1.
Consider \((S',H') \in \mathcal{S}\) with precisely one child \((S,H)\) in $D$.
We first claim that  \(H \neq \emptyset\) and \((S',H') = (S \cup H,\emptyset)\).
Since \((S,H)\) is a child of \((S',H')\) in \(D\), we have \((S,H) < (S',H')\).
Also note that for every \(x \in S' \setminus (S \cup H)\)
we have \((\{x\},\emptyset) < (S',H')\) and \((\{x\},\emptyset) \not \leq (S,H)\). Hence, \((S',H')\) has outdegree at least~2 in \(D\), a contradiction.
Thus, we have \(S' \subseteq S \cup H\) and, in view of the definition of \(<\), this implies 
\(S \cup H = S'\). Now, if there exists some \(x \in H'\)
then \((\{x\},\emptyset) < (S',H')\) and \((\{x\},\emptyset) \not \leq (S,H)\), then it follows
again that \((S',H')\) has outdegree at least~2 in \(D\), a contradiction.
Thus we must have \(H' = \emptyset\), implying, in view of \((S,H) \neq (S',H')\),
that \(H \neq \emptyset\), which completes the proof of the claim.

Now, consider the directed path
\[P(S,H) = ((S_*,H_*),(S_k,H),(S_{k-1},H),\dots,(S_1,H),(H,\emptyset))\]
and note that we have \((S_*,H_*) = (S',H')\). Since \((S',H')\) has
precisely one child in $D$, Proposition~\ref{prop:char:reticulation:cycle:structure}
implies that \(P(S,H)\) is not contained in any reticulation cycle in~$D$.
Thus, in the construction of \(\mathcal{N}(\mathcal{S})\) the arc
from \((S',H') = (S_*,H_*)\) to \((H,\emptyset)\) is added. Hence, \((S',H')\)
has outdegree at least~2 in \(N\). It follows that there is no vertex in \(N\) with outdegree~1.

Next note that every vertex of \(N\) has indegree at most~2,
since by Proposition~\ref{prop:char:indegree:at:most:two}, every vertex of~$D$
has indegree
at most~2, and we only add arcs in the construction of \(N\) from~$D$
whose end vertex has indegree~1 in~$D$.

Finally,  we show that no two distinct
reticulation cycles in \(N\) have an arc in common.
By Proposition~\ref{prop:char:reticulation:cycle:structure},
every reticulation cycle \(\mathcal{C}\) in \(N\) is either a reticulation cycle in $D$
or it arises by adding an arc from the start vertex to the end vertex of the directed path \(P(S,H)\) in $D$
for some \((S,H) \in \mathcal{S}\) with \(H \neq \emptyset\)
for which \(P(S,H)\) is not already contained in a reticulation cycle in $D$.
But then, again in view of Proposition~\ref{prop:char:reticulation:cycle:structure}, no two distinct
reticulation cycles in \(N\) can have an arc in common.
\qed

\noindent We now prove the main result of this section.\\

\noindent{\em Proof of Theorem~\ref{char:theo:encodings:nc}:}
Consider a set pair system \(\mathcal{S}\) on \(X\). 
As noted at the beginning of this section,
by Proposition~\ref{proposition:network:set:pair:systems}, if \(\mathcal{S} = \theta(\mathcal{N})\)
for some compressed 1-nested network \(\mathcal{N}\) on \(X\), then \(\mathcal{S}\) is 1-nested compatible.

Conversely, assume that \(\mathcal{S}\) is a 1-nested compatible set pair
system on \(X\). Then, by Theorem~\ref{theorem:char:ns:is:1:nested},
\(\mathcal{N}(\mathcal{S})\) is a compressed 1-nested network on \(X\).
To show that \(\theta(\mathcal{N}(\mathcal{S})) = \mathcal{S}\),
it suffices to show that for every vertex \(u=(S,H)\) of
\(N(\mathcal{S}) = ((\mathcal{S},A'),\varphi)\) we have \(\theta(u) = (S,H)\).
To this end, first note that for every \(x \in S \cup H\), we have, by the definition of \(\leq\),
\((\{x\},\emptyset) \leq (S,H)\). Similarly, for every \(x \not \in S \cup H\)
we have \((\{x\},\emptyset) \not \leq (S,H)\).
Thus, in view of \(\varphi(x) = (\{x\},\emptyset)\) for all \(x \in X\),
we have \(S(u) \cup H(u) = S \cup H\). Hence, to show that
\(S(u) = S\) and \(H(u) = H\) it remains to establish that \(H(u) = H\).
We consider two cases.

\emph{Case 1}: \(H = \emptyset\).
Assume for contradiction that \(H(u) \neq \emptyset\).
Then there must exist some \((S_1,H_1) \in \mathcal{S}\)
with \((S_1,H_1) < (S,H)\) such that \((S_1,H_1)\) is a child of some
\((S_2,H_2) \in \mathcal{S}\) with \((S_2,H_2) \not \leq (S,H)\).
This implies that \((S_1,H_1)\) has indegree~2 and, thus,
\((S_1,H_1)\) is the end vertex of the two paths in a
reticulation cycle \(\mathcal{C}\) in \(\mathcal{N}(\mathcal{S})\).
Hence, we have \(H_1 = \emptyset\) and, in view of \(\emptyset \neq S_1 \subseteq S \cap (S_2 \cup H_2)\),
(NC4) implies \((S,H) < (S_2,H_2)\).
So, \((S_2,H_2)\) must be the start vertex of the two directed paths in \(\mathcal{C}\)
and \((S,H)\) is a vertex on one of these directed paths distinct from the start
vertex and the end vertex. But this implies \(H \neq \emptyset\), a contradiction.

\emph{Case 2}: \(H \neq \emptyset\).
By the construction of \(\mathcal{N}(\mathcal{S})\),
\((S,H)\) is a vertex other than the start vertex and the end vertex in
the directed path \(P(S,H)\) and
this directed path is contained in a reticulation cycle \(\mathcal{C}\) in \(\mathcal{N}(\mathcal{S})\).
The end vertex of the two paths in \(\mathcal{C}\) is \(v = (H,\emptyset)\).
Thus, we have \(H(u) = S(v)\) and, by Case 1, we have \(S(v) = H\),
implying \(H(u) = H\).

Now, having established that \(\theta(\mathcal{N}(\mathcal{S})) = \mathcal{S}\),
we complete the proof of Theorem~\ref{char:theo:encodings:nc} 
by noting that in view of Theorem~\ref{theorem:theta:is:encoding}, every
compressed 1-nested network \(\mathcal{N}\) on \(X\) with
\(\theta(\mathcal{N}) = \mathcal{S}\) is isomorphic 
to \(\mathcal{N}(\mathcal{S})\). \qed

\section{Consensus networks}
\label{section:computing:consensus}

In this section we present a way to derive a consensus network for a non-empty
collection \(\mathfrak{C}\) of compressed 1-nested networks on~\(X\).
To this end, for such a collection \(\mathfrak{C}\), put
\(\theta(\mathfrak{C}) = \bigcup_{\mathcal{N} \in \mathfrak{C}} \theta(\mathcal{N})\)
and denote, for every \((S,H) \in \theta(\mathfrak{C})\),
by \(\#(S,H)\) the number of networks \(\mathcal{N} \in \mathfrak{C}\)
with \((S,H) \in \theta(\mathcal{N})\). In addition, for real numbers \(p\) and \(q\) with
\(0 \leq p < 1\) and \(0 \leq q < 1\), put \(\theta(\mathfrak{C})_{(p,q)}\) to be the
set pair system
\[\{(S,H) \in \theta(\mathfrak{C}) : \#(S,H) > p|\mathfrak{C}| \ \text{and} \ H = \emptyset\}
  \cup \{(S,H) \in \theta(\mathfrak{C}) : \#(S,H) > q|\mathfrak{C}| \ \text{and} \ H \neq \emptyset\}.\]

\noindent Our main result in this section is then as follows:

\begin{theo}
\label{theorem:computing:consensus}
Given a collection \(\mathfrak{C}\) of \(t \ge 1\) compressed 1-nested
networks on a set \(X\) with \(|X| = n\),  
\(\mathcal{N}(\theta(\mathfrak{C})_{(\frac{1}{2},\frac{2}{3})})\) 
is a compressed 1-nested consensus network of \(\mathfrak{C}\)
that can be computed in \(O(tn^2 + n^3)\) time.
\end{theo}

Note that if all networks in~\(\mathfrak{C}\) are phylogenetic trees
(so that \(H = \emptyset\) holds for all \((S,H) \in \theta(\mathfrak{C})\)),
then  \(\mathcal{N}(\theta(\mathfrak{C})_{(\frac{1}{2},\frac{2}{3})})\)  is
the majority rule consensus tree mentioned in the introduction.

To prove Theorem~\ref{theorem:computing:consensus}, we first consider 
for which choices of $p$ and $q$ the set 
\(\theta(\mathfrak{C})_{(p,q)}\) is 1-nested compatible.

\begin{lem}
\label{lemma:majority:rule:consensus:works}
For all \(\frac{1}{2} \leq p < 1\) and all \(\frac{2}{3} \leq q < 1\), the set pair system
\(\theta(\mathfrak{C})_{(p,q)}\) is 1-nested compatible.
\end{lem}

\pf
Clearly, \((X,\emptyset)\) and \((\{x\},\emptyset)\) for all \(x \in X\) are
contained in \(\theta(\mathcal{N})\) for all \(\mathcal{N} \in \mathfrak{C}\)
and so they are contained in \(\theta(\mathfrak{C})_{(p,q)}\).
Thus \(\theta(\mathfrak{C})_{(p,q)}\) satisfies (NC1) and (NC2).

If \((S,H)\) with \(H \neq \emptyset\) is contained in \(\theta(\mathfrak{C})_{(p,q)}\)
then there exist more than \(\frac{2}{3}|\mathfrak{C}|\) networks \(\mathcal{N} \in \mathfrak{C}\) with
\((S,H) \in \theta(\mathcal{N})\). Since \(\theta(\mathcal{N})\) satisfies (NC3), for all
these networks we also have \((H,\emptyset) \in \theta(\mathcal{N})\), implying that
\((H,\emptyset) \in \theta(\mathfrak{C})_{(p,q)}\). Hence (NC3) holds for \(\theta(\mathfrak{C})_{(p,q)}\).

To establish that (NC4) holds for \(\theta(\mathfrak{C})_{(p,q)}\),
consider any two distinct \((S,H)\), \((S',H') \in \theta(\mathfrak{C})_{(p,q)}\).
Since both \(\#(S,H) > \frac{1}{2}|\mathfrak{C}|\) and \(\#(S',H') > \frac{1}{2}|\mathfrak{C}|\), by the pigeon hole
principle, there must exist some  \(\mathcal{N} \in \mathfrak{C}\) with
\((S,H)\), \((S',H') \in \theta(\mathcal{N})\), implying that
precisely one of the conditions given in (NC4) holds for
\((S,H)\) and \((S',H')\) since \(\theta(\mathcal{N})\) is 1-nested compatible.

Finally, to establish that also (NC5) holds for \(\theta(\mathfrak{C})_{(p,q)}\),
consider any three distinct \((S_1,H_1)\), \((S_2,H_2)\), \((S_3,H_3) \in \theta(\mathfrak{C})_{(p,q)}\).
Since \(\#(S_i,H_i) > \frac{2}{3}|\mathfrak{C}|\) for all \(i \in \{1,2,3\}\), by the pigeon hole
principle, there must exist some  \(\mathcal{N} \in \mathfrak{C}\) with
\((S_1,H_1)\), \((S_2,H_2)\), \((S_3,H_3) \in \theta(\mathcal{N})\),
implying that the condition given in (NC5) holds for
\((S_1,H_1)\), \((S_2,H_2)\) and \((S_3,H_3)\) since \(\theta(\mathcal{N})\) is 1-nested compatible.
\qed

\begin{figure}
\centering
\includegraphics[scale=0.9]{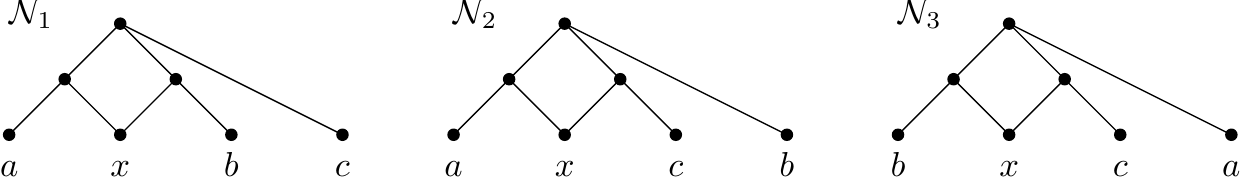}
\caption{For the collection \(\mathfrak{C} = \{\mathcal{N}_1,\mathcal{N}_2,\mathcal{N}_3\}\)
of compressed 1-nested networks on \(X=\{a,b,c,x\}\)
the set pair system \(\theta(\mathfrak{C})_{(\frac{1}{2},q)}\) is not 1-nested compatible
for any \(0 \leq q < \frac{2}{3}\).}
\label{figure:example:consensus:constants}
\end{figure}

Note that the lower bounds of \(\frac{1}{2}\) for \(p\) and
\(\frac{2}{3}\) for \(q\) in Lemma~\ref{lemma:majority:rule:consensus:works}
are the smallest possible (cf. Figure~\ref{figure:example:consensus:constants}).
This also implies that some condition involving three set pairs such as (NC5)
cannot be avoided in any characterization of those set pair systems that arise
as encodings of isomorphism classes of compressed 1-nested networks.

To derive an upper bound on the run time for computing 
\(\mathcal{N}(\theta(\mathfrak{C})_{(\frac{1}{2},\frac{2}{3})})\), we
rely on an upper bound for the  
size of a 1-nested compatible set pair system. In
view of Theorem~\ref{char:theo:encodings:nc} finding such a bound 
is equivalent to giving an upper bound on the number of vertices in a compressed
1-nested network on \(X\) in terms of \(n=|X|\).  
In view of upper bounds on the number of vertices in the
closely related level-1 networks given e.g. in~\cite[Lemma~4.5]{van2009algorithms}
and~\cite[Lemma~3.1]{gambette2017challenge}, the following result
is perhaps not surprising, however we give its proof for the sake of completeness:

\begin{lem}
\label{lemma:size:one:nested:compatible:set:pair:system}
Let \(\mathcal{S}\) be a 1-nested compatible set pair system
on a set \(X\) with \(|X| = n\). Then \(|\mathcal{S}| \leq 3n-2\)
and this upper bound is tight.
\end{lem}

\pf
As mentioned above, it suffices to
consider an arbitrary compressed 1-nested network \(\mathcal{N}\) on a set \(X\) with \(|X| = n\)
and to establish that \(|\theta(\mathcal{N})| \leq 3n-2\).
Also note that if \(\mathcal{N}\) does not contain any reticulation cycle then \(\mathcal{N}\) is a rooted
phylogenetic tree on \(X\) and it is known that \(|\theta(\mathcal{N})| \leq 2n - 1\)
(see e.g. \cite[Sec.~2.1]{sem-ste-03a}).

So assume that \(\mathcal{N}\) contains at least one reticulation cycle \(\mathcal{C} = \{P,P'\}\).
Without loss of generality we assume that the directed path \(P\) consists of at least three vertices.
Let \(e = (u,v)\) be the last arc on \(P\). Note that \(u\) has indegree~1
in \(\mathcal{N}\). We remove \(e\) from \(\mathcal{N}\). If after the removal of \(e\)
vertex \(u\) has outdegree~1 we suppress~\(u\). We perform this removal of an arc
for every reticulation cycle in \(\mathcal{N}\) and obtain a rooted tree \(\mathcal{T}\)
on \(X\) with
\[|\theta(\mathcal{N})| \leq |\theta(\mathcal{T})| + c(\mathcal{N}) \leq 2n - 1 + c(\mathcal{N}),\]
where \(c(\mathcal{N})\) is the number of reticulation cycles in \(\mathcal{N}\).

Now, to establish \(|\theta(\mathcal{N})| \leq 3n-2\), it suffices to show
that \(c(\mathcal{N}) \leq n-1\) by induction on \(n\). The base case of
the induction for \(n=2\) claims that any compressed 1-nested network with
precisely two leaves contains at most~1 reticulation cycle,
which can easily be checked to be true. For \(n \geq 3\), consider the root
\(\rho\) of \(\mathcal{N}\). To apply the induction hypothesis, we split
\(\mathcal{N}\) at \(\rho\) into two networks \(\mathcal{N}_1\) and \(\mathcal{N}_2\)
on disjoint non-empty subsets \(X_1\) and \(X_2\) of \(X\) with \(X_1 \cup X_2 = X\).
Note that if \(\rho\) has outdegree~2 and is contained in a reticulation cycle
this involves the removal of an arc from this reticulation cycle as described
in the previous paragraph. By induction, we have
\[c(\mathcal{N}) \leq c(\mathcal{N}_1) + c(\mathcal{N}_2) + 1 \leq (|X_1|-1) + (|X_2|-1) + 1 = n-1,\]
as required.

\begin{figure}
\centering
\includegraphics[scale=0.9]{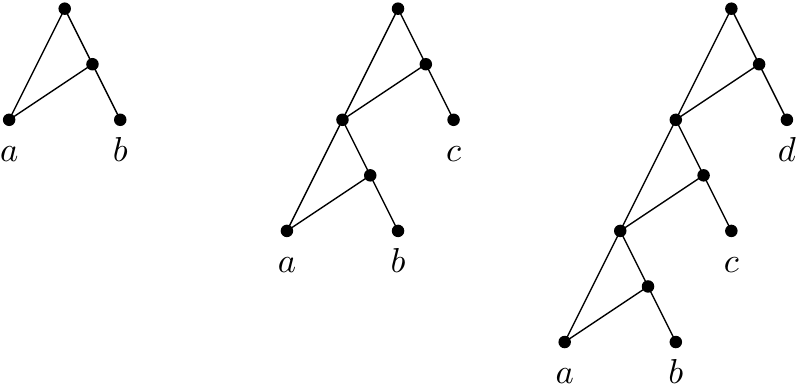}
\caption{Examples of compressed 1-nested networks \(\mathcal{N}\)
         on a set \(X\) with \(|X| = n\) and \(|\theta(\mathcal{N})| = 3n - 2\)
         for \(n \in \{2,3,4\}\).}
\label{figure:example:upper:bound:size:compatible:set:pair:system}
\end{figure}

It remains to note that, for every \(n \geq 2\), there exists a compressed 1-nested network
\(\mathcal{N}\) on a set \(X\) with \(|X| = n\) and \(|\theta(\mathcal{N})| = 3n-2\).
In Figure~\ref{figure:example:upper:bound:size:compatible:set:pair:system}
examples for \(n \in \{2,3,4\}\) are depicted that can easily be generalized to
any \(n \geq 5\).
\qed

\noindent We now prove the main result of this section.\\
 
\noindent{\em Proof of Theorem~\ref{theorem:computing:consensus}:}
By Lemma~\ref{lemma:majority:rule:consensus:works}, the set pair system
\(\mathcal{S}=\theta(\mathfrak{C})_{(\frac{1}{2},\frac{2}{3})}\) is 1-nested compatible. Thus,
in view of Theorem~\ref{char:theo:encodings:nc}, we have the compressed
1-nested network \(\mathcal{N} = \mathcal{N}(\mathcal{S})\) on \(X\) 
with \(\theta(\mathcal{N}) = \mathcal{S}\).

To bound the run time for computing \(\mathcal{N}\), note that for
every \(\mathcal{N}' \in \mathfrak{C}\) the set pair system
\(\theta(\mathcal{N}')\) can be computed in \(O(n^2)\) time
since \(\theta(\mathcal{N}')\) is 1-nested compatible and
therefore \(|\theta(\mathcal{N}')| \in O(n)\) by Lemma~\ref{lemma:size:one:nested:compatible:set:pair:system}.
Next, using a trie as the data structure for storing set pairs, \(\theta(\mathfrak{C})\)
together with the number \(\#(S,H)\) for all \((S,H) \in \theta(\mathfrak{C})\)
can be computed in \(O(tn^2)\) time. Note that \(|\theta(\mathfrak{C})| \in O(tn)\).
Then, \(\mathcal{S} = \theta(\mathfrak{C})_{(\frac{1}{2},\frac{2}{3})}\) can be computed
in \(O(tn + n^2)\) in view of the fact that \(\mathcal{S}\) is 1-nested compatible
and, therefore, \(|\mathcal{S}| \in O(n)\) again by Lemma~\ref{lemma:size:one:nested:compatible:set:pair:system}.
Finally, performing a transitive reduction~\cite{aho-gar-72a} on the DAG corresponding to the
partial order \(\leq\) on \(\mathcal{S}\), we obtain the Hasse diagram \(D(\mathcal{S})\)
in \(O(n^3)\) time and from \(D(\mathcal{S})\) we get the compressed
1-nested network \(\mathcal{N}(\mathcal{S})\) in \(O(n)\) time.
Overall, this yields a run time in \(O(tn^2 + n^3)\).
\qed

Before concluding this section we
note that as a consequence of Lemma~\ref{lemma:size:one:nested:compatible:set:pair:system} 
we can also give a bound on the time complexity of 
checking whether or not a set pair system is 1-nested compatible.

\begin{cor}
\label{corollary:complexity:decide:1:nested}
Let \(\mathcal{S}\) be a set pair system with \(|\mathcal{S}| = k\) on a set \(X\) with \(|X| = n\).
Then it can be checked in \(O(k + n^3)\) time whether or not \(\mathcal{S}\) is 1-nested compatible.
\end{cor}

\pf
We first compute \(k = |\mathcal{S}|\). By Lemma~\ref{lemma:size:one:nested:compatible:set:pair:system},
if \(k > 3n-2\) then \(\mathcal{S}\) cannot be 1-nested compatible and we are done. Otherwise
we need to check if conditions (NC1)-(NC5) hold for \(\mathcal{S}\).
For (NC1)-(NC4) this can be done in \(O(n^3)\) time by directly checking the conditions.
For (NC5) we first partition \(\mathcal{S}\) in \(O(n^2)\) time by putting
\[\mathcal{H} = \{H : (S,H) \in \mathcal{S}\}\]
and then computing 
\[\mathcal{S}(H) = \{(S',H') \in \mathcal{S} : H' = H\}\]
for every \(H \in \mathcal{H}\).
Then, for all \(H \in \mathcal H - \{\emptyset\}\) and 
for all \((S',H), (S'',H) \in \mathcal{S}(H)\), we precompute
whether or not each of the following holds: \(S' \cap S'' = \emptyset\), \(S' \subsetneq S''\),
\(S'' \subsetneq S'\). This preprocessing can be done in \(O(n^3)\) time.
Then, for any three distinct \((S_1,H), (S_2,H), (S_3,H) \in \mathcal{S}(H)\) we can compute
in constant time whether (i)~\(S_1 \cap S_2 = \emptyset\), \(S_1 \subsetneq S_3\) and \(S_2 \subsetneq S_3\) holds
as well as whether (ii)~\(S_1 \cap S_2 = S_1 \cap S_3 = S_2 \cap S_3 = \emptyset\) holds.
Note that checking (i) and (ii) is equivalent to checking (NC5) and,
thus, (NC5) can also be checked in \(O(n^3)\) time.
\qed

\section{Discussion} \label{section:discussion}

We have presented a new characterization of an encoding of
compressed 1-nested networks
and used it to develop a novel approach to compute a consensus
for a collection of such networks. These results open up 
various new directions and lead to several 
questions  including the following (see~\cite[Chapter 10]{steel2016phylogeny} 
for an overview of phylogenetic networks and the definitions for the classes that we mention):

\begin{itemize}
\item
Can similar encodings be given and characterized for other classes of phylogenetic networks?
For example, in~\cite{cardona2008comparison} an encoding 
for so-called {\em tree-child networks} is presented, and it would be 
interesting to understand how these encodings can be characterized. Other 
classes of phylogenetic networks that could be interesting to 
consider in this context are \emph{level}-$k$ networks for small $k \geq 2$, \emph{normal} networks and 
\emph{unrooted} phylogenetic networks. 

\item
The majority rule consensus tree can be unresolved in practice, and approaches such
as the loose and greedy consensus are used to deal with this issue~\cite{bryant2003classification}.
Can  such techniques be developed for our approach?  For example, a 1-nested 
compatible set pair system on \(X\) could also be 
computed greedily from \(\theta(\mathfrak{C})\) for~\(\mathfrak{C}\)
a collection of compressed 1-nested networks.
Once some 1-nested compatible set pair system \(\mathcal{S}\) on \(X\) 
has been computed from \(\theta(\mathfrak{C})\),
by Theorem~\ref{char:theo:encodings:nc},
\(\mathcal{N}(\mathcal{S})\) yields a consensus network of~\(\mathfrak{C}\).

\item
Is it possible to improve the run time in Theorem~\ref{theorem:computing:consensus}?
One approach to addressing this question may be to use ideas 
similar to those presented in~\cite{bra-cor-09a} to compute
\(\mathcal{N}(\mathcal{S})\) in \(O(n^2)\) time. 	
	
\item
There are several alternatives to using the majority rule for computing the consensus
of a collection of phylogenetic trees~\cite{bryant2003classification}. Can
any of these be also extended to 1-nested networks? 
For example, another approach to encoding 1-nested networks 
given in~\cite{huber2013encoding} uses 3-leaved subnetworks 
called {\em trinets}; these are also used to encode level-2 and tree-child networks 
in~\cite{van2014trinets}. Can consensus methods for phylogenetic trees using 
triplets (e.g. the local consensus tree described in~\cite[p.8]{bryant2003classification}) 
be extended to 1-nested networks using trinets and, if so, what are their mathematical properties?
	
\item
Can axiomatic properties of consensus
methods for 1-nested networks be developed along the lines of those for
phylogenetic trees described in~\cite{day2003axiomatic}
(see also~\cite[Chapter 2.6]{steel2016phylogeny})?
\end{itemize}

\section*{Acknowledgements}
Huber and Moulton thank the Lorentz Center, Leiden, The Netherlands,
and the organizers of the workshop ``Distinguishability in
Genealogical Phylogenetic Networks" 
held at that center in 2018, where they were inspired to start
thinking about some 
of the ideas presented in this paper.

\bibliographystyle{abbrvurl}
\bibliography{encoding_one_nested_networks}

\end{document}